\documentclass[twocolumn,useAMS,usenatbib]{mnras} \usepackage{natbib}

\usepackage{amsmath,latexsym,amssymb} \usepackage{epsfig,graphicx} \topmargin-1cm
\usepackage{makecell}
\usepackage{epstopdf}
\usepackage{float}
\usepackage{fixltx2e}
\usepackage{color}
\usepackage{url}
\graphicspath{{figures/},{images_JWST/},{quasar_properties/},{newfigs/}}
\DeclareGraphicsExtensions{.pdf,.png,.jpg,.eps}
\def\reff@jnl#1{{\rm#1\/}}

\def\aj{\reff@jnl{AJ}}                  % Astronomical Journal
\def\araa{\reff@jnl{ARA\&A}}            % Annual Review of Astron and Astrophys
\def\apj{\reff@jnl{ApJ}}                        % Astrophysical Journal
\def\apjl{\reff@jnl{ApJ}}               % Astrophysical Journal, Letters
\def\apjs{\reff@jnl{ApJS}}              % Astrophysical Journal, Supplement
\def\apss{\reff@jnl{Ap\&SS}}            % Astrophysics and Space Science
\def\aap{\reff@jnl{A\&A}}               % Astronomy and Astrophysics
\def\aapr{\reff@jnl{A\&A~Rev.}}         % Astronomy and Astrophysics Reviews
\def\aaps{\reff@jnl{A\&AS}}             % Astronomy and Astrophysics, Supplement
\def\baas{\reff@jnl{BAAS}}              % Bulletin of the AAS
\def\jcap{\reff@jnl{JCAP}}              % Journal of Cosmology and Astroparticle Physics
\def\jrasc{\reff@jnl{JRASC}}            % Journal of the RAS of Canada
\def\memras{\reff@jnl{MmRAS}}           % Memoirs of the RAS
\def\mnras{\reff@jnl{MNRAS}}            % Monthly Notices of the RAS
\def\physrep{\reff@jnl{Phys.Rep.}}
\def\pra{\reff@jnl{Phys.Rev.A}}         % Physical Review A: General Physics
\def\prb{\reff@jnl{Phys.Rev.B}}         % Physical Review B: Solid State
\def\prc{\reff@jnl{Phys.Rev.C}}         % Physical Review C
\def\prd{\reff@jnl{Phys.Rev.D}}         % Physical Review D
\def\prl{\reff@jnl{Phys.Rev.Lett}}      % Physical Review Letters
\def\pasp{\reff@jnl{PASP}}              % Publications of the ASP
\def\pasj{\reff@jnl{PASJ}}              % Publications of the ASJ
\def\skytel{\reff@jnl{S\&T}}            % Sky and Telescope
\def\solphys{\reff@jnl{Solar~Phys.}}    % Solar Physics
\def\sovast{\reff@jnl{Soviet~Ast.}}     % Soviet Astronomy
\def\ssr{\reff@jnl{Space~Sci.Rev.}}     % Space Science Reviews
\def\nat{\reff@jnl{Nature}}             % Nature

\newcommand{\hmpc}{\ensuremath{h^{-1}\mathrm{Mpc}}}

\newcommand{\Msun}{M_{\odot}}

\newcommand{\beq}{\begin{equation}}
\newcommand{\eeq}{\end{equation}}
\newcommand{\beqa}{\begin{eqnarray}}
\newcommand{\eeqa}{\end{eqnarray}}

\title[BT-II Quasar]{A tiny host galaxy for the first giant black hole:  $z= 7.5$ quasar in BlueTides} 
\author[A. Tenneti et al.]{\parbox{18cm} {Ananth
  Tenneti$^{1}$\thanks{\tt vat@andrew.cmu.edu}, Stephen M. Wilkins $^2$\thanks{\tt S.Wilkins@sussex.ac.uk},  Tiziana Di
  Matteo$^1$\thanks{\tt tiziana@phys.cmu.edu},
  Rupert A.C. Croft$^1$
  and Yu Feng$^3$ }
  \vspace{0.3cm}
  \\$^1$ McWilliams Center for
  Cosmology, Department of Physics, Carnegie Mellon University,
  Pittsburgh, PA 15213, USA\\
$^2$ Department of Physics and Astronomy, University of Sussex, Brighton BN1 9QH, UK\\
$^3$ Berkeley Center for Cosmological Physics, Department of Physics, University of California Berkeley, Berkeley, CA 94720, USA}

\date{\today}
\begin{document}
\maketitle

\begin{abstract}
The most distant known quasar recently discovered by \cite{2018Natur.553..473B} is at $z=7.5$ (690 Myr after the Big Bang), at the dawn of galaxy formation. We explore the host galaxy of the brightest quasar in the large volume cosmological hydrodynamic simulation BlueTides, which in Phase II has reached these redshifts. The brightest quasar in BlueTides has a luminosity of a $\sim$ few $10^{13} L_{\odot}$ and a black hole mass of $6.4 \times 10^{8} M_{\odot}$ at $z \sim 7.5$, comparable to the observed quasar (the only one in this large volume). The quasar resides in a rare halo of mass $M_{H} \sim 10^{12} M_{\odot}$ and has a host galaxy  of stellar mass of $4 \times 10^{10}M_{\odot}$ with an ongoing (intrinsic) star formation rate of $\sim 80 M_{\odot} yr^{-1}$. The corresponding intrinsic UV magnitude of the galaxy is $-23.1$, which is roughly $2.7$ magnitudes fainter than the quasar's magnitude of $-25.9$. We find that the galaxy is highly metal enriched with a mean metallicity equal to the solar value.
We derive quasar and galaxy spectral energy distribution (SED) in the mid and near infrared JWST bands. We predict a significant amount of dust attenuation in the rest-frame UV
corresponding to $A_{1500} \sim 1.7$ giving an UV based SFR of
$\sim 14 M_{\odot} yr^{-1}$.
We present mock JWST images of the galaxy with and without central point source, in different MIRI and NIRCam filters. The host galaxy is detectable in NIRCam filters, but it is  extremely compact ($R_{E}=0.35$ kpc). It will require 
JWST's exquisite sensitivity and resolution to separate the galaxy
from the central point source.  Finally within the FOV of the quasar in BlueTides there are two more sources that would be detectable by JWST.
\end{abstract}


\section{Introduction} \label{S:intro}

There are many unsolved problems in our quest to understand the first billion years of cosmic structure formation and the formation of the first galaxies and black holes. Supermassive black holes, as massive as those in galaxies
today, are known to exist in the early universe, even up to $z\sim 7$. 
Luminous, extremely rare, quasars at $z \sim 6$ have been discovered in the Sloan Digital Sky Survey \citep{{2006AJ....132..117F},{2009AJ....138..305J}} and, until recently, the highest
redshift quasar known was \citep{2015Natur.518..512W} at $z = 7.09$ \citep{2011Natur.474..616M}. 

Excitingly, \cite{2018Natur.553..473B} (hereafter B18) reported the discovery of a bright quasar at $z=7.54$, J1342 + 0928 which is currently the record holder for known high redshift quasars. The observed quasar is found to have a bolometric luminosity of $4 \times 10^{13} L_{\odot}$ and an inferred black hole mass of $8 \times 10^{8} M_{\odot}$. However, the properties of the galaxy hosting this quasar are currently completely unknown. In this paper, we focus on predicting the host galaxy properties of such a luminous and massive early quasar using the cosmological hydrodynamic simulation BlueTides \cite{2016MNRAS.455.2778F}. 
More specifically, we make predictions for the Spectral Energy Distributions (SED's) of the quasar host galaxy and the AGN in the James Webb Space Telescope (JWST) \citep{2006SSRv..123..485G} filters.  In a companion paper \citep{2018arXiv.xx000x}, we study the feedback around the most luminous quasar, and in particular, the properties of the outflow gas in the host halo of the quasar.  
    
The upcoming launch of JWST will facilitate the observational study of the properties of a large number of high redshift galaxies at $z \sim 8-10$. In particular, JWST will make observations in the rest-frame optical/near-IR wavelengths of high redshift galaxies including the massive quasar host galaxies. \cite{2017ApJ...849..155V} predicted the properties of high-redshift galaxies and AGN in JWST bands by developing a population synthesis model based on empirical relations. The volume and resolution of BlueTides simulation facilitates a direct study of AGN and host galaxy properties of rare objects such as most massive quasars at high redshift ($z \sim 7.5$). In this paper, we compare the SED's of galaxies which are directly calculated based on the stellar age and metallicity that are then compared with the quasar's SED obtained based on luminosity and accretion rate.
 
 Recently, \cite{2017ApJ...851L...8V} reported observations from IRAM/NOEMA and JVLA to obtain some constraints on the host galaxy of the quasar, J1342 + 0928 of B18.  The  dynamical mass of the host is determined to be $< 1.5 \times 10^{11} M_{\odot}$ and star formation rates are in the range, $85-545 M_{\odot} yr^{-1}$. \cite{2017ApJ...851L...8V} also report a dust mass of $0.6-4.3 \times 10^{8}M_{\odot}$, a metal enriched gas ($\sim 5 \times 10^{6} M_{\odot} M_{C^{+}}$) with implied stellar mass of $\sim 2 \times 10^{10} M_{\odot}$   
 
This paper is organized as follows. In Section~\ref{S:Simulation}, we provide the details of the Bluetides-II simulation along with the methods to identify galaxies and compute spectral energy distributions (SED's). In Section~\ref{galaxy_prop}, we provide details of the properties of the most luminous quasar and it's host galaxy in BT-II. We show the distribution of gas, dark matter and stellar matter in BT-II corresponding to a region of JWST field-of-view (FOV) in Section~\ref{S:checkfov}. In Section~\ref{S:seds}, we discuss the AGN and host galaxy SED'S along with their band luminosities in the JWST filters. The mock JWST images of the host galaxy, sampled at JWST resolution and including PSF effects of the filters are shown in Section~\ref{galimages}. Finally, we conclude in Section~\ref{S:conclusions}.

\section{Bluetides-II Simulation} \label{S:Simulation}
 
The BlueTides simulation \citep{2016MNRAS.455.2778F} was the first phase in the BlueTides project, and involved the evolution of a cosmological volume to $z=8$. 
In this paper we report on some of the first results from BlueTides-II (BT-II), the second phase of the project, which continued the evolution of the BlueTides
volume to redshifts $z<8$. Here we focus on redshifts $z=7.5-8$, close to the redshift of the J1342$+$0928 quasar in B18.

\subsection{MP-Gadget SPH code}

Both phases of the BlueTides simulation have been performed using the Smoothed Particle Hydrodynamics code, MP-Gadget\footnote{\url{https://github.com/MP-Gadget/MP-Gadget}} \citep{2016MNRAS.455.2778F} in a cubical periodic box of volume $(400 \hmpc)^{3}$. The simulation was evolved from initial conditions at $z=99$ with $2 \times 7040^{3}$ dark matter and gas particles. The cosmological parameters in the simulation were chosen according to those in the  WMAP 9 year data release \citep{2013ApJS..208...19H}. The details of phase I of the  simulation and various properties of the simulated galaxies till $z=8$ are described in \cite{2016MNRAS.455.2778F}.

We provide a brief description of the feedback model adopted in the BlueTides simulation. The MP-Gadget code adopts the pressure-entropy formulation of smoothed particle hydrodynamics (pSPH) \citep{{2010MNRAS.405.1513R},{2013MNRAS.428.2840H}} to solve the Euler equations. Star formation is implemented based on the multi-phase star formation model \citep{2003MNRAS.339..289S} and also includes several modifications following \cite{2013MNRAS.436.3031V}. The gas cooling is modeled based on radiative processes \citep{1996ApJS..105...19K} and also via metal cooling\citep{2014MNRAS.444.1518V}. The formation of molecular hydrogen and its effect on star formation at low metallicities is modeled according to the prescription by \cite{2011ApJ...729...36K}. A type II supernovae wind feedback model \citep{2010MNRAS.406..208O}, which assumes that the wind speeds are proportional to the local one dimensional dark matter velocity dispersion is also incorporated. Finally, the black hole growth and feedback from active galactic nuclei (AGN) is incorporated based on the super-massive black hole model developed in \cite{2005Natur.433..604D}.

\begin{figure}
\includegraphics[width=\columnwidth]{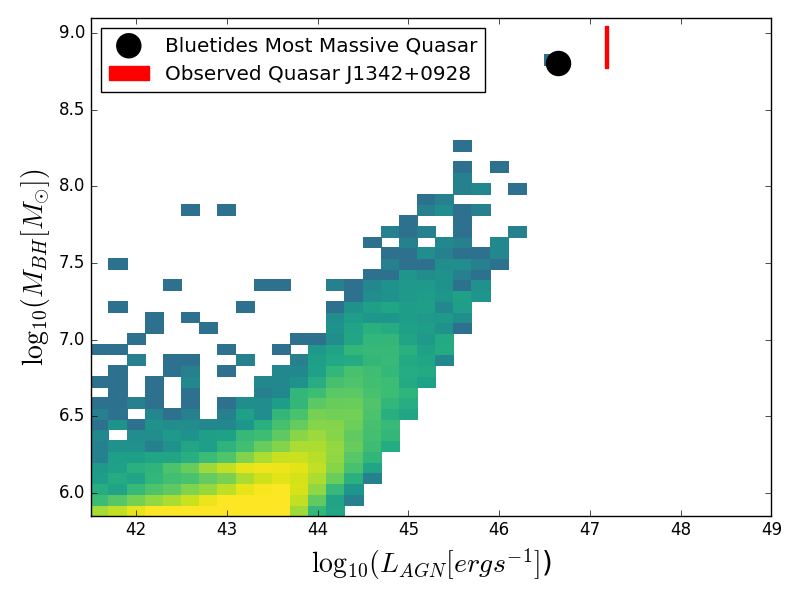}
\caption{\label{fig:lagn_mbh} The bolometric  luminosity of AGN in the BT-II simulation at redshift $z=7.6$ plotted against black hole
mass. The solid black circle is plotted at the position of the most luminous
quasar in BT-II, which we study in detail in this paper, along with its host galaxy. The red line shows the relevant information for the observed $z=7.54$ quasar, J1342$+$0928 of B18. The height of the line corresponds to the observational uncertainty. }
\end{figure}

\subsection{Galaxies}

We identify galaxies from the snapshots of BT-II using a friends of
friends (FOF) algorithm \citep{1985ApJ...292..371D} with linking length 0.2 times the mean interparticle separation. We have shown in \cite{2016MNRAS.455.2778F} that at these redshifts, there is a good correspondence between FOF defined objects and galaxies  selected using sExtractor \citep{1996A&AS..117..393B} from mock imaging. In the simulation at
redshift $z=7.5$  there are $4.7\times 10^{5}$ galaxies with stellar mass $>$ $5 \times 10^{7} M_{\odot}$.

Each galaxy contains from $60 - 2.3\times10^{5}$ star particles, each with an age and metallicity.
We use the information from these to compute spectra for the galaxies, and
also for the visualization of galaxy images in different bands. 

\subsection{Galaxies SEDs}\label{galseds}
The spectral energy distribution of the AGN host is constructed by attaching the SED of simple stellar population (SSP) to each star particle based on its mass, age, and, metallicity. 
Specifically we employ version 2.1 of the Binary Population and Spectral Populations Synthesis (SPS) \citep[BPASS][]{2017PASA...34...58E} model utilizing a modified Salpeter IMF (Salpeter high-mass slope with a break at $<0.5 M_{\odot}$) and a high-mass cut-off of $300. M_{\odot}$. See \citep{2016MNRAS.460.3170W} for a discussion of the impact of assuming alternative SPS models. Attenuation by dust is modeled for each star particle individually by determining the line-of-density of metals and assuming a linear relation to the dust optical depth. This is calibrated to reproduce the bright end of the rest-frame UV luminosity function at $z\sim8 $\citep[see][]{2017MNRAS.469.2517W}.

\subsection{Quasar spectrum}\label{agnseds}
The spectral energy distribution of the AGN is determined by the black hole mass and accretion rate using the functional form of the spectrum adopted in the spectral synthesis code, {\em Cloudy } \citep{2013RMxAA..49..137F}, given by 
\begin{equation} \label{eq1}
f_{\nu} = \nu ^{\alpha _{UV}}e ^{-\frac{h \nu}{k T_{BB}}} e ^{- \frac{k T_{IR}}{h \nu}} + a \nu ^{\alpha _{x}}
\end{equation}

where $\alpha _{UV} = -0.5$, $\alpha _{X} = -1$, $k T_{IR} = 0.01Ryd$. We note that this procedure to obtain the theoretical AGN spectrum is the same as the method used in \cite{2017ApJ...849..155V}.

\subsection{The most luminous quasar in BlueTides-II}

The quasar, J1342$+$0928 found by B18 was discovered using data that
cover a significant fraction of the sky. 
Three large area surveys were used, the Wide-field Infrared Survey Explorer
(ALLWISE), the United Kingdom
Infrared Telescope Infrared Deep Sky Survey (UKIDSS) Large Area Survey, and the DECam Legacy
Survey (DECaLS)\footnote{\url{http://legacysurvey.org/decamls}}. The overlap between these
is up to 4000 square degrees, which leads to a potential search volume
of up to 5.5 (Gpc/h)$^{3}$ between $z=7.5$ and $z=8$. Although the BlueTides volume
is large for a cosmological hydrodynamic simulation, the observational search
volume could be up to two orders of magnitude larger. This should be kept 
in mind when comparing the simulation and observational results.
When searching for a luminous quasar in the simulation, however, we do allow ourselves to search over a slightly wider redshift range than the single
snapshot $z=7.54$ corresponding to the observed quasar's redshift. Because
of the variability in black hole accretion, and therefore quasar luminosity,
by picking a redshift where the luminosity is high we are 
increasing the effective simulation volume.

\begin{figure}
\includegraphics[width=\columnwidth]{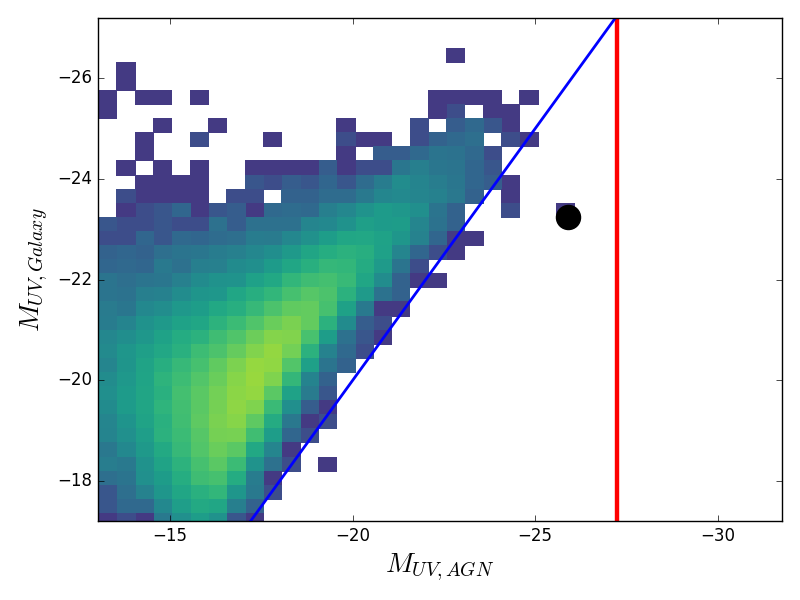}
\caption{\label{fig:Lgalaxy_Lagn_MAG} The absolute magnitude of AGN in the BTII simulation at redshift $z=7.6$ plotted against galaxy absolute
magnitude. The solid black circle is plotted at the position of the most luminous
quasar in BT-II. The red line shows the observed luminosity of 
J1342$+$0928 quasar in B18.The solid blue line shows where galaxy and AGN are equally luminous. The host galaxy is predicted to have $M_{uv}= -23.24$ }
\end{figure} 

\begin{figure}
%\vspace{-1cm}
\includegraphics[width=\columnwidth]{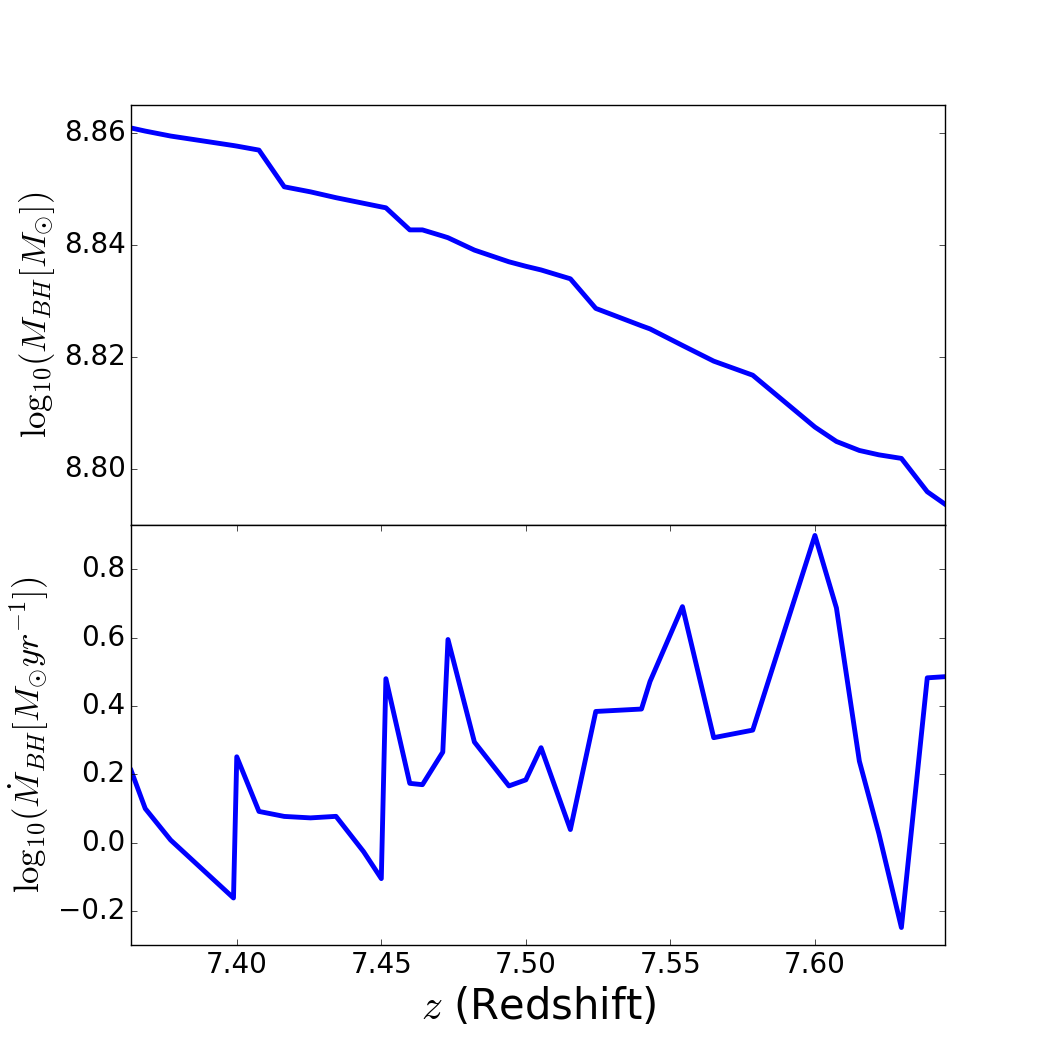}
%\vspace{-3cm}
\caption{\label{F:fig_quasar_props}  Redshift evolution of the black hole mass ({\em Top}), and black hole accretion rate.}
\end{figure}

We choose the object in BT-II to compare to the B18 quasar, J1342$+$0928 on the
basis of its luminosity. In Figure~\ref{fig:lagn_mbh} we show black hole mass against AGN luminosity for the entire simulation at redshift $z=7.6$. We can 
see that there is a clear relationship between the two quantities, and also that the brightest AGN also has the most massive black hole. The black hole mass,
accretion rate and other parameters related to the host galaxy are given in 
Table \ref{T:quasar_properties}. The black hole mass is within the 1$\sigma$  observational uncertainty quoted by B18, and its luminosity is smaller by a factor of $\sim 3.4$.  We study this black hole and its host galaxy in the rest
of the paper.

Over the redshift range $z=7.5-8$ which we consider, the black hole is growing,
as can be seen in Figure~\ref{F:fig_quasar_props}. The accretion rate is between
$1-10 \Msun$/yr, with variations of a factor of 2 or 3 over timescales of $\sim 50$ Myr.
These variations are large compared to the observational error on the B18 quasar 
luminosity, indicating that there is no need for the simulation quasar
luminosity to be an exact match.
The simulated black hole's accretion rate does appear to be slowing down over the range plotted, which
can also be seen as the flatter trend of black hole growth in the top panel.
This is consistent with the quasar's surroundings having been affected by 
AGN feedback. We study this feedback in more detail for the same simulated
object in a companion paper \citep{2018arXiv.xx000x}.

\section{Quasar and galaxy properties}\label{galaxy_prop}

\begin{figure}
\begin{center}
\vspace{-1cm}
\includegraphics[width=\columnwidth]{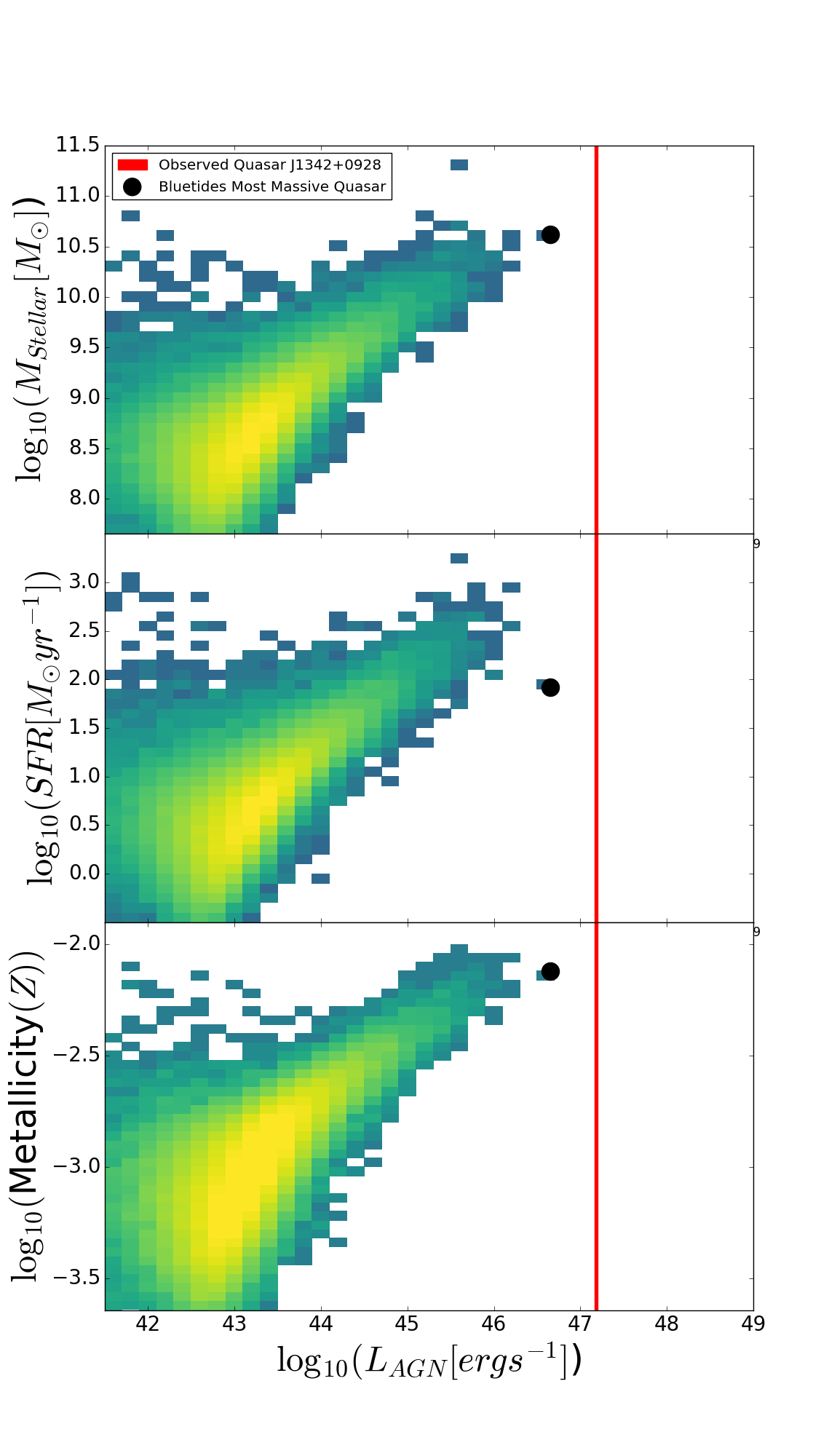}
\vspace{-0.5cm}
\caption{\label{F:fig_gal_properties} Properties of the host 
galaxy for the brightest quasar.
{\em From Top to Bottom:} 
The stellar mass, $M_\odot$,
the star formation rate (SFR) 
and stellar metallicities 
of the galaxies in BTII
at $z=7.6$. The most massive/luminous object is indicated by a solid black point in previous figures.}  
\end{center}
\end{figure}

The most luminous AGN in BT-II at $z\sim 7.5$ is hosted by a galaxy with FoF halo mass $\sim 10^{12}\Msun$, and stellar mass $4 \times 10^{10} \Msun$
(see Table \ref{T:quasar_properties}). It is however not the most massive
galaxy in the volume, as can be seen from the top panel of Figure~\ref{F:fig_gal_properties}, where we show the stellar mass of all galaxies against AGN luminosity. There are about 10 other galaxies with stellar masses
which are  similar or greater, including some with AGN luminosities 5
orders of magnitude smaller. The most massive galaxy has a stellar mass $\sim$ 5 times
larger than the host of the most luminous AGN. 

When considering galaxy luminosity, the quasar host is even less exceptional,
which can be seen in Figure~\ref{fig:Lgalaxy_Lagn_MAG}. 
Here we plot the absolute restframe UV magnitude of the AGN and their host galaxies. We can see that the host of brightest AGN is about 3 magnitudes fainter than the brightest galaxies in BT-II. Only a handful of galaxies lie to the right of the $M_{UV,AGN}=M_{UV,Galaxy}$ line,
indicating that the vast majority of accreting black holes do not outshine their hosts. The main locus of
points in Figure \ref{fig:Lgalaxy_Lagn_MAG} is approximately
3 magnitudes to the left of (fainter than) the $M_{UV,AGN}=M_{UV,Galaxy}$ line, indicating that for most galaxies the AGN is 3 magnitudes fainter the galaxy. The brightest quasar is the opposite, on the other hand, being about 3 magnitudes more luminous than its host.

\begin{figure}
\vspace{-0.11cm}
\includegraphics[width=\columnwidth]{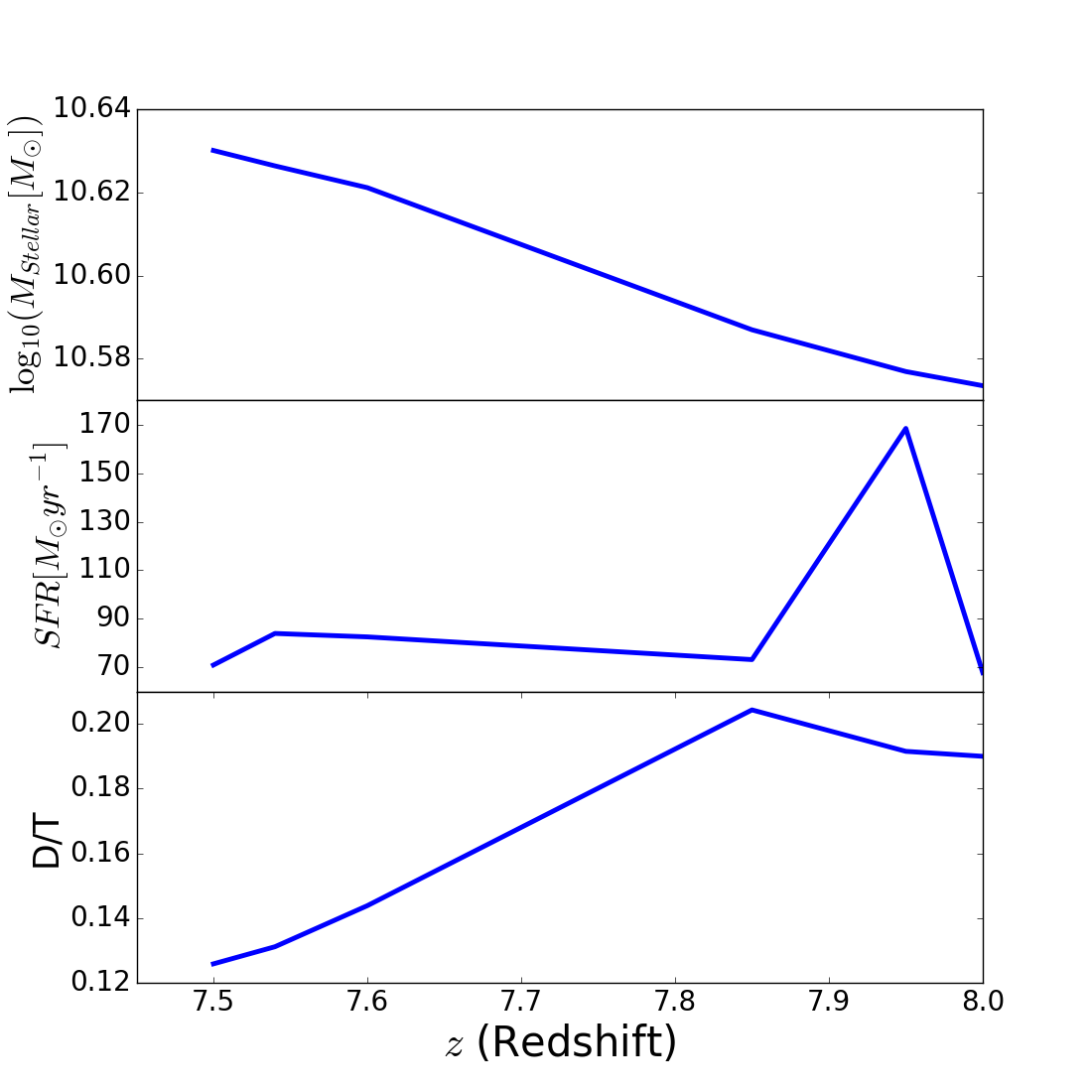}
\vspace{-0.5cm}
\caption{\label{F:fig_gal_properties1} The host galaxy's stellar mass ({\em  Top}), star formation rate ({\em Middle}) and the Disk to Total ratio ({\em Bottom}) as a function of redshift.}  
\end{figure}

The galaxy UV luminosity is closely related to the star formation rate.
The star formation rate of the host galaxy is computed from the sum of the SFRs of all particles in the halo. 
From the middle panel of \ref{F:fig_gal_properties} we can see that the 
star formation rate of the brightest quasar host is about 1 order of magnitude smaller than the most
star forming. The star formation history of this host galaxy does  include episodes of much higher activity. For example, in Figure 3 of \cite{2017MNRAS.467.4243D}, the same galaxy is shown at much higher redshifts,
where at $z=10$, the star formation rate was as high as $300 \Msun$/yr. Even at
redshifts closer (within $\Delta z \sim 0.5$) to the one we are considering,
the star formation rate was an order of magnitude higher, as we see below
(Figure~\ref{F:fig_gal_properties1}).

Figure~\ref{F:fig_gal_properties} also shows the stellar mass of the galaxies as a function of AGN
luminosity. In this case, the AGN host is again in the
top few galaxies, consistent with the fact that it underwent a
very significant 
earlier burst of star formation. This star formation
episode resulted in metal production, which is evident
in the bottom panel of Figure~\ref{F:fig_gal_properties}, where it is also among the
top few galaxies by mean metallicity (computed from
the star particles). We note that the solar metallicity is defined to be $Z_{\odot}=0.02$ in the simulation.

We show the SFR as function of redshift 
over the interval of interest in the middle panel
of Figure~\ref{F:fig_gal_properties1}.
The sampling is relatively coarse in time because of the timing of the simulation snapshots, but we can see that it has varied by a factor of around two over the 150 Myr before $z=7.5$. The growth in the stellar mass of the galaxy is shown in the top panel. 

We have also computed the galaxy's disk to total ratio using star particle kinematics. We
have used a standard technique \citep{2007MNRAS.374.1479G} to determine the fraction of stars in each galaxy that are on planar
circular orbits and that are associated with a bulge.
In \cite{2017MNRAS.467.4243D} (where more details are given) we found that the majority of massive galaxies at
these high redshifts are disks.
In fact, at $z=8$,  70\% of galaxies above a mass of
$10^{10}\Msun$ were kinematically classified as disks (using the standard
threshold of disk stars to total stars (D/T) ratio of 0.2 \citep{2007MNRAS.374.1479G}. In the present case, the quasar host galaxy is below this threshold, and is becoming even less disk-like towards lower redshifts. 
Some of the galaxies in \cite{2017MNRAS.467.4243D} were extremely disk-like in visual appearance as well as kinematically. As we shall see later (Section~\ref{galimages_native}), this galaxy is not. This is consistent with the general expectation for  galaxies hosting the most massive black holes which are often in areas with low tidal fields \cite{2017MNRAS.467.4243D}, and are less likely to be disks.

\begin{figure}
\vspace{-1.3cm}
\includegraphics[width=1.2\columnwidth]{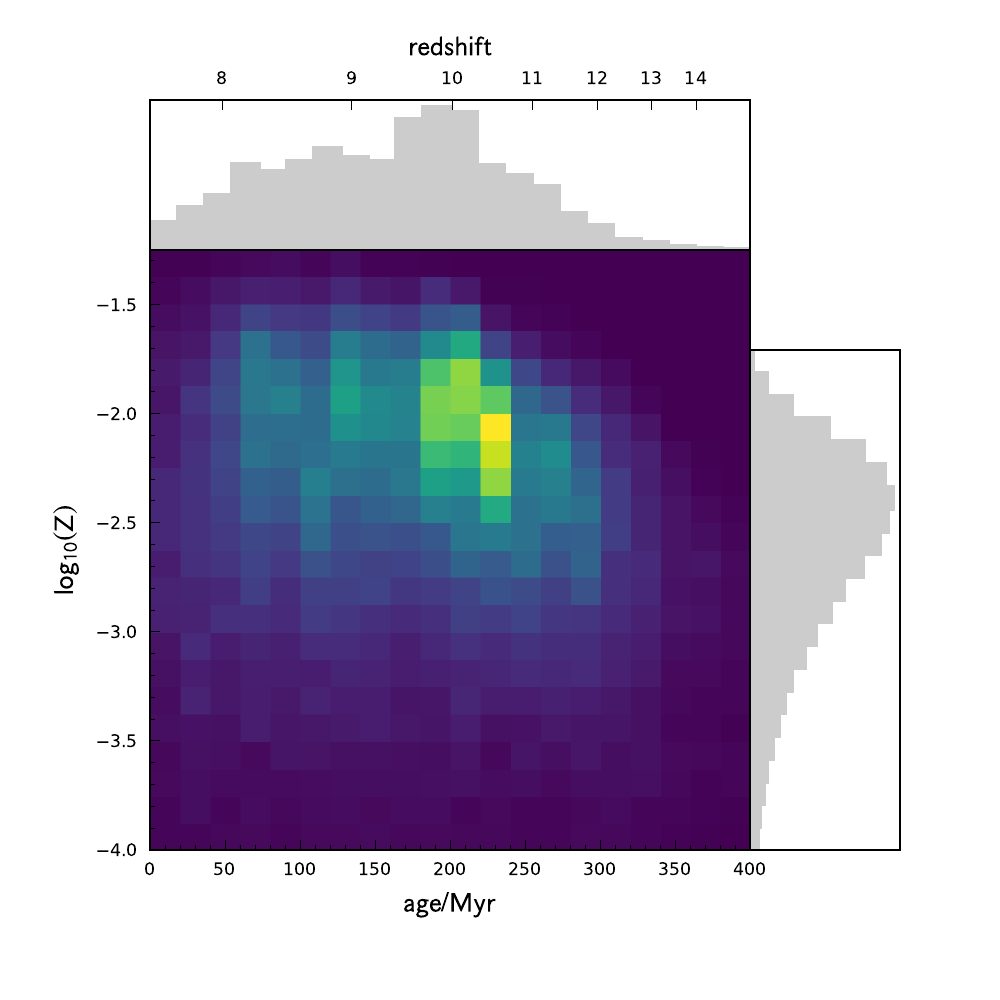}
\vspace{-0.5cm}
\caption{\label{metalage}The metallicity of star particles in the quasar host galaxy as a function of their age (or equivalently formation redshift). Solar metallicity is $Z=0.02$.}
\end{figure}

With the large number of particles in the quasar host galaxy, the metallicity of the stars as a function of  age can be examined in detail. We do this in Figure~\ref{metalage}, where we show a two dimensional histogram of these properties. The median metallicity of the stars that formed at redshifts $z>12$ can be seen to be about $log_{10}(Z)\simeq -2.5$ and this increases to 
log$_{10}(Z)\simeq -1.75$  by redshift $z=8$. As expected
for these early sites of vigorous star formation \citep[e.g.,][]{{2016ApJ...816...37V},{2013ApJ...773...44W},{2011AJ....142..101W},{2005A&A...440L..51M},{2003Natur.424..406W}}, the median metallicity is 
high, reaching solar values for stars forming at $z\sim9$.
The burst of star formation at $z\sim10$ in this galaxy can be clearly seen in the age histogram above the main plot, and also as a concentration of pixels in the main panel at these redshifts and metallicity $\log _{10}(Z)\simeq -2.1$.

\section{Large scale environment: the JWST field-of-view (FOV)} \label{S:checkfov}
  
The most luminous quasar in BT-II is located in a high density environment which will make it an interesting target for observations of the surrounding sky area. We are most interested in follow-up observations with the JWST, and in this section we present images of
the dark matter, gas and stellar in the entire JWST field of view, before concentrating on mock images of the host galaxy in detail in Section~\ref{galimages}. 

\begin{table}
\caption{\label{T:quasar_properties} }
\centering\begin{tabular}{ccccc}
\hline
 Property & z=8 & z=7.85 & z=7.6 & z=7.54\\
\hline
$M_{BH}$ ($\times 10^{8} 
M_{\odot}$) & 4.1 & 5.2 & 6.4 & 6.7 \\
$D/T$ & 0.19 & 0.20 & 0.14 & 0.13 \\
$L_{BH}$ $(\times 10^{12} L_{\odot})$ & 5.7 & 5.9 & 11.8 & 3.6 \\
$\dot{M}_{BH}$ $(M_{\odot}/yr)$ & 3.8 & 4.0 & 7.9 & 2.5 \\
$SFR$ $(M_{\odot}/yr)$ & 67.80 & 73.19 & 82.55 & 83.96 \\
\makecell{$M_{Halo}$ \\ ($\times 10^{11} h^{-}M_{\odot}$)} & 5.9 & 6.1 & 9.1 & 9.4 \\
\makecell{$M_{200}$ \\ ($\times 10^{11} h^{-1}M_{\odot}$)} & 3.7 & 3.9 & 4.3 & 4.4 \\
\makecell{$M_{star}$ \\ ($\times 10^{10} M_{\odot}$)} & 3.75 & 3.86 & 4.18 & 4.23 \\
\makecell{$M_{gas}$ \\ ($\times 10^{10} h^{-1} M_{\odot}$)} & 4.9 & 4.8 & 9.3 & 9.5 \\
\makecell{$f_{gas}$ \\$\left(\frac{M_{gas}}{M_{gas} + M_{star}}\right)$} & 0.65 & 0.64 & 0.76 & 0.76 \\
\hline
\end{tabular}
\end{table}

JWST's Mid-Infrared Instrument (MIRI) and Near Infrared Camera (NIRCam) have different FOV sizes. The  NIRCam instrument has 2 modules each with a FOV of $132^{``} \times 132^{``}$ with filters in the wavelength range, $0.6 - 5.0 \mu m$. MIRI has a FOV of $74^{``} \times 113{``}$ and provides broadband imaging in the wavelength range, $5.0 - 27.0 \mu m$.  The physical scale corresponding to $1^{''}$ at $z=7.6$ is $5.1 kpc$, so that the NIRCam and MIRI FOV's are approximately $4  \hmpc$ across their longest dimensions. We therefore choose to plot images of a $4  \hmpc \times 4  \hmpc$ 
simulation. We also restrict ourselves to a depth of $ 4 \hmpc$ to
only show physically associated structure. The volume considered is
therefore $10^{-6}$ of the entire BT-II simulation box.

\begin{figure*}
\vspace{-2cm}
\hbox{
\hspace{-2cm}
\includegraphics[width=4.1in]{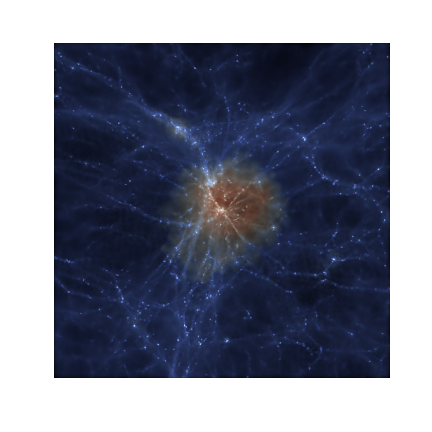}
\hspace{-1cm}
\includegraphics[width=4.1in]{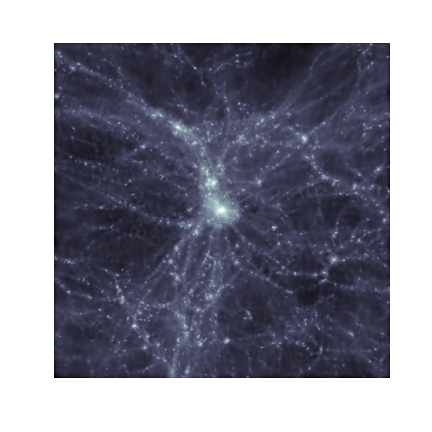} } 
\vspace{-1cm}
\caption{\label{F:fig_fov} Images showing the distribution of gas and  dark matter in the region centered at the most massive black hole of size $4\hmpc$ corresponding to the JWST field of view at $z=7.6$. {\em Top Left:} Distribution of gas where the intensity of the blue region represents the density and the color scale represents the temperature of the gas. {\em Top Right}: the corresponding dark matter density.}
\end{figure*}

\subsection{Gas and dark matter}  
  
In Figure~\ref{F:fig_fov} (left panel), we plot the distribution of gas in this region of $4\hmpc$ centered on the most massive black hole in the BT-II simulation. The filamentary structure of the gas is readily visible, with the quasar itself lying at the intersection of four filaments. The distribution of matter is visually fairly symmetric around the center, indicating that the asymmetry in the tidal field will be fairly low, consistent with the low D/T ratio
of the host galaxy \cite{2017MNRAS.467.4243D}. The gas in the image is color coded by temperature, with the IGM far from the galaxy being at a temperature
of $\sim 10^{4}$ K. The BT-II simulation includes a patchy model for reionization, but such a high density region has already been reionized (heating the IGM to this temperature) by this redshift. The
red-orange color of gas close to the quasar indicates that it has been heated by AGN feedback to temperatures of $10^{6}-10^{7}$K. This feedback and the associated winds are studied in more detail in \cite{2018arXiv.xx000x}.

The dark matter distribution around the quasar  (Figure~\ref{F:fig_fov}, left panel), traces the gas distribution, as expected. A large filament is more visible in dark matter than gas, entering from the top left, and in general throughout the image many subhalos are visible along the filaments. 

\begin{figure*}

\includegraphics[width=5.4in]{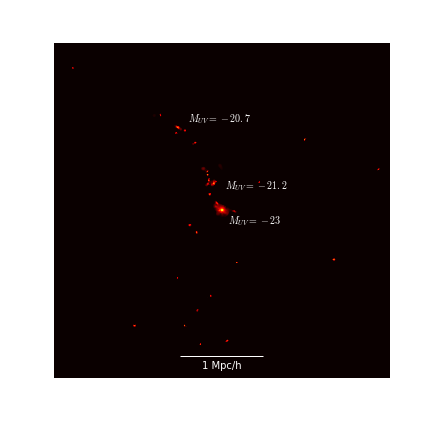}
\vspace{-1.2cm}
\caption{\label{F:fig_fov_star} Galaxies in the field of
view of JWST. 
Stellar densities are shown  increasing from  red to yellow. The labels indicate the absolute magnitudes of the brightest three galaxies in the restframe UV.}  
\end{figure*}

\subsection{Stars} 

An interesting question we would like to address is whether any companion galaxies are expected close to the location of the highest redshift quasar. Deep JWST imaging may be able to detect close neighbors or even satellite galaxies. 
In Figure~\ref{F:fig_fov_star}, the intensity of the red region represents the stellar density. We also label the rest frame UV magnitudes of three most luminous galaxies in the region. Here, the galaxy with the most massive black hole has the highest luminosity in this field of view with a $M_{UV}=-23$. We can see two more luminous galaxies, one slightly above this galaxy with a magnitude of $M_{UV} = -21.2$ and another one further above at the left with a magnitude of $M_{UV}=-20.7$. The second brightest galaxy is within 300 kpc of the
quasar in projected separation and the host halos seen in Figure 
are close to merging.

\begin{figure*}
\begin{center}
\includegraphics[width=3.4in]{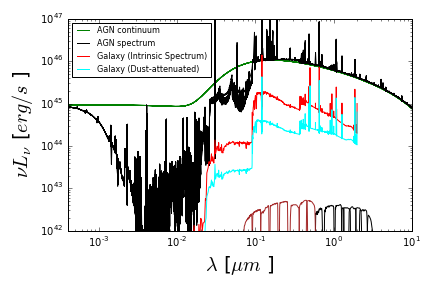}
\includegraphics[width=3.4in]{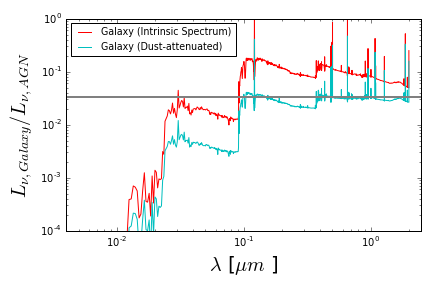}
\caption{\label{F:fig_sed} {\em Left:} The figure shows the SED's of AGN and the star particles in the galaxy hosting the brightest quasar at $z=7.6$ . The green and black lines show the continuum and emission spectrum of the quasar obtained from the {\em Cloudy} software. The red line shows the galaxy's intrinsic spectrum obtained by summing the individual SED's of star particles in the galaxy. The cyan line shows the dust-attenuated spectrum of the galaxy {\em Right:} The figure shows the ratio of the galaxy luminosity to that of the AGN as a function of wavelength. The red and cyan line correspond to the intrinsic and dust-attenuated spectrum of the galaxy respectively. The gray horizontal line represents the value at the which the AGN luminosity is about 30 times brighter than the galaxy. 
}  
\end{center}
\end{figure*}

If the extinction for the other galaxies is assumed to be the same as for quasar host galaxy, then their luminosities will be fainter by $1.5$ magnitudes for each galaxy .  The extincted UV magnitudes of the galaxies labeled in the figure will therefore be $-21.5$,$-19.7$, $-19.2$ respectively in the decreasing order of their luminosities.
All three of the galaxies visible and labeled in Figure~\ref{F:fig_fov} (right panel)  would be visible in JWST imaging, for example using JWST's NIRCam F070W filter (see below) and an apparent magnitude limit of  $m_{AB} \sim28$ for an integration time\footnote{\url{https://jwst-docs.stsci.edu/display/JTI/NIRCam+Sensitivity}} of $10\mathrm{ks}$. The galaxy with the most massive black hole has apparent magnitude of $m_{AB} = 24.1$. The other two labeled galaxies visible in this FOV have magnitudes of $m_{AB} = 26.9$ and $m_{AB} = 27.5$.

\section{Quasar and Galaxy SED's} \label{S:seds}

We determine the luminosities of the quasar host galaxy in each JWST MIRI and NIRCam filters by summing the Spectral Energy Distributions (SED's) of each star particle computed from the method described in Section~\ref{galseds}  and convolved with the given filter.
The SED of the quasar is obtained using the spectral synthesis code, {\em Cloudy} \citep{2013RMxAA..49..137F} based on the black hole mass and accretion rate as described in Section~\ref{agnseds}. 
The rest frame quasar spectrum is shown in the left panel of Figure~\ref{F:fig_sed}. Here, we show both the AGN continuum (green lines) and the emission spectrum (black line). The filter responses of the different mid and near infrared JWST bands in the quasar rest frame are also shown at the bottom of the plot. We can see that the peak of the AGN SED is captured by the near infrared filters, as is the Lyman break and Lyman-alpha absorption.

\begin{figure*}
\includegraphics[width=1.3\columnwidth]{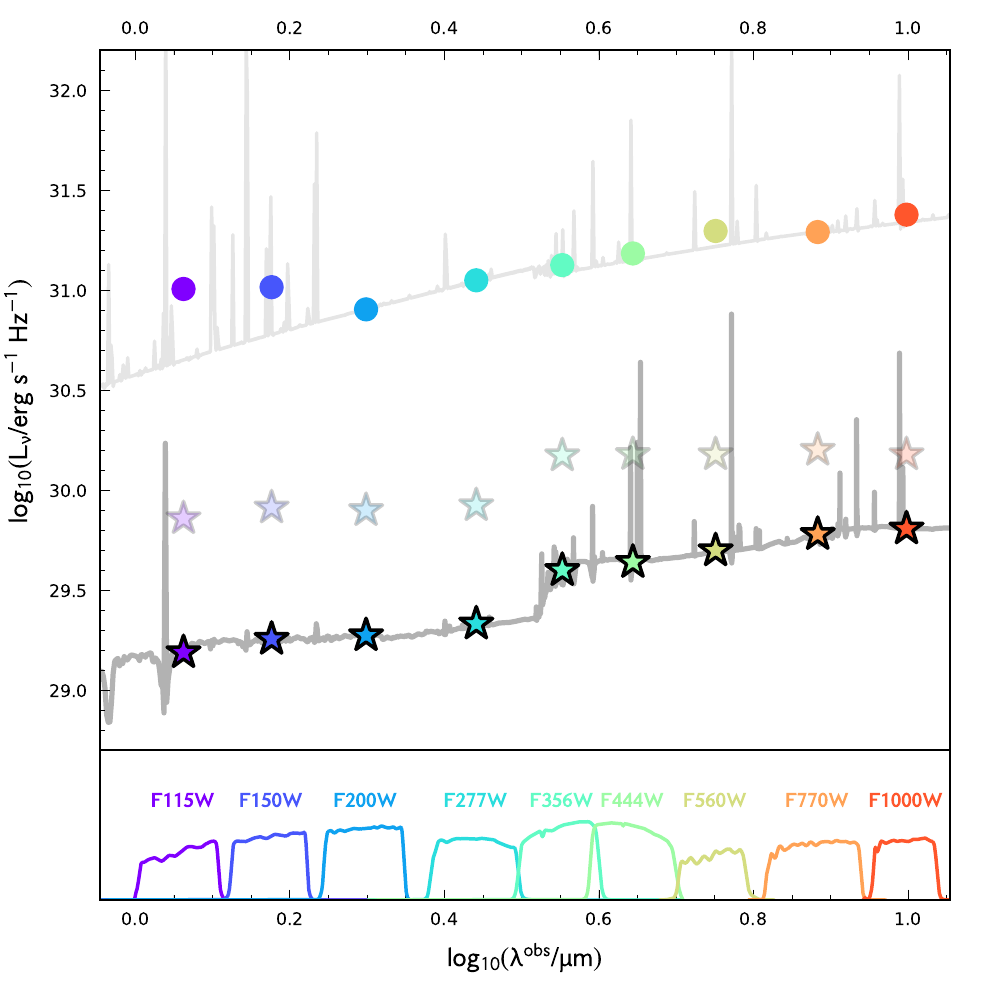}
\caption{\label{F:fig_sed_obs} The {\em Top panel} in the figure shows the observed-frame spectra of the brightest quasar (gray line) and the host galaxy (black line). The {\em Bottom panel} shows the filter curves of JWST's NIRCam F115W, F150W, F200W, F277W, F356W, F444W filters and MIRI F560W, F770W, F1000W filters. The corresponding band luminosities of the quasar (Circular markers) and the host galaxy (Star markers) are shown at the mean wavelengths in the {\em top} panel.}  
\end{figure*}

The AGN outshines the host galaxy considerably in all JWST wavebands. In the left panel of Figure~\ref{F:fig_sed} the SED of the quasar is compared against the galaxy spectrum obtained by summing the individual SED's of star particles in the galaxy. 
We show both the galaxy's intrinsic spectrum (red line) and the dust-attenuated spectrum (cyan line). 
To understand the relative brightness of the host galaxy when compared to that of its quasar at different wavelengths, we plot the ratio of the SED's of the galaxy and the quasar in the right panel of Figure~\ref{F:fig_sed}. 
As seen from the figure, at rest frame wavelengths above $0.5 \mu m$, the  dust attenuated luminosity of the galaxy is smaller than that of the quasar by up a factor of $\sim 30$. More specifically
the dust attenuation in the rest-frame UV gives an extinction $A_{1500} \sim 1.7$
and, compared to the SFR based on the intrinsic stellar UV luminosity
of  SFR$_{int}(L_ {1500}) \sim 65 \Msun$/yr, the resulting 
SFR based on the attenuated stellar UV luminosity SFR$_{obs}(L_{1500}) \sim 14 \Msun$/yr.

\begin{figure*}
\includegraphics[width=2.1\columnwidth]{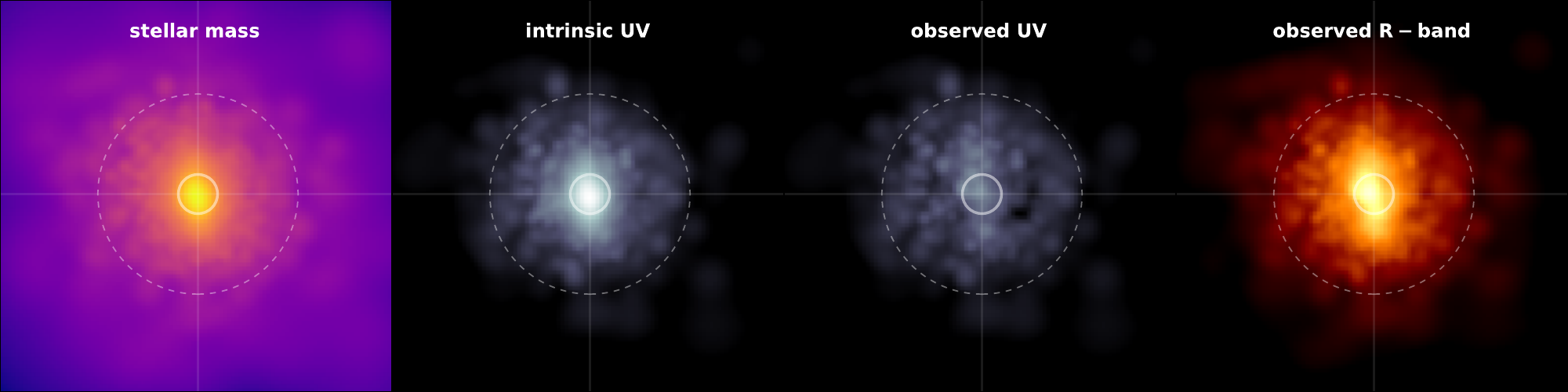}
\caption{\label{F:fig_mock_stellarmass} Images of the host galaxy of brightest quasar in a region of size $10 kpc$ in physical units with a resolution of $0.1 kpc$. The inner circle is of radius $1 kpc (0.2^{"})$ and the outer circle depicts a radius of $5 kpc (1^{"})$. From Left to Right, the images show the distribution of stellar mass, intrinsic UV-band luminosity, observed UV-band luminosity and observed R-band luminosity}
\end{figure*}

We focus in more detail in the NIR part of the spectrum in Figure \ref{F:fig_sed_obs}, where we plot the observed frame monochromatic luminosities of the quasar and galaxy as a function of wavelength. 
At observed wavelengths above $3 \mu m$, (NIRCam F356W, F444W filters, MIRI F560W, F770W, F1000W filters) the luminosities are separated by $\sim 1.5 dex$. Point source subtraction techniques are slightly
more likely to be successful at long wavelengths. Based on the 
galaxy-AGN luminosity ratio, this is likely to be challenging if
the host galaxy of the observed quasar, J1342$+$0928 in B18 is similar to the equivalent host
in the simulation.

\section{Mock JWST Images of host galaxy}\label{galimages}

With an angular resolution of $\sim 0.1^{"}$, JWST will be able to 
image the host galaxy of J1342 + 0928. The BT-II simulation has a
resolution (force softening) of 250 pc at this redshift, which 
corresponds to $0.05^{"}$, so the simulation is well matched
to the potential of JWST, and can be used to predict how the galaxy
will appear in JWST imaging.
Here, we first show native images (without convolution with telescope angular or spectral characteristics) and them move on to mock JWST images of the individual host galaxy of the bright quasar at $z=7.6$. 

\subsection{Native images without PSF convolution and filter selection}\label{galimages_native}

Before plotting the images of the galaxy in different JWST filters, we show the distribution of only the stellar component of the galaxy in the left panel of Figure~\ref{F:fig_mock_stellarmass}. 
The image shows a region of width, $10 kpc$ in physical units. The inner circle in the images has a radius of $1 kpc$, while the outer circle is of radius $5 kpc$, corresponding to a size of $1^{``}$.
We note that the galaxy is centrally concentrated, but it is visible out to $\sim 8 kpc$. It is ellipsoidal and rather featureless. We show only one orientation, but the other views are similar - it exhibits no flattening or disk-like structure.

\begin{figure*}
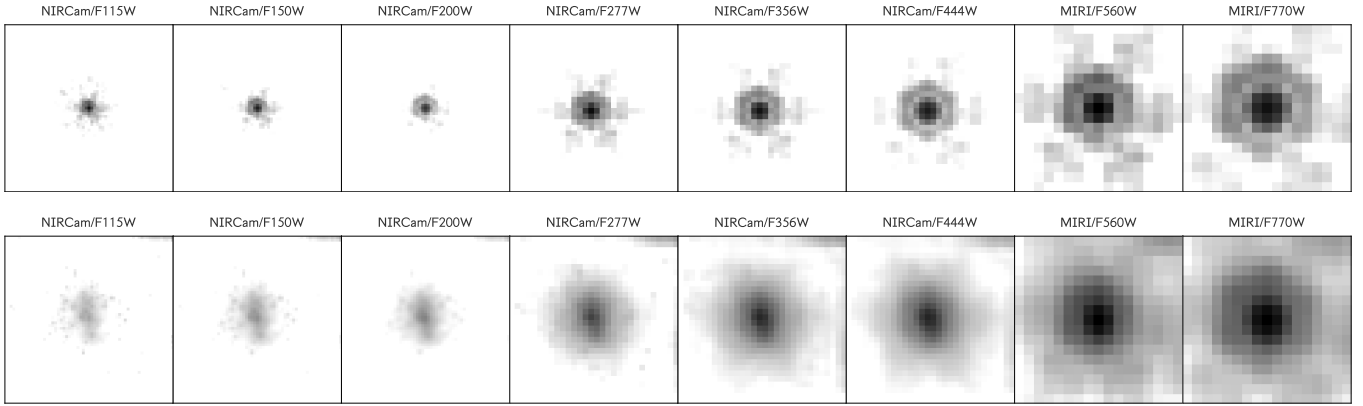

\vbox{
\includegraphics[width=2.1\columnwidth]{JWST_default_AGN_logged.pdf} \\
\includegraphics[width=2.1\columnwidth]{JWST_default_noAGN_logged.pdf}}
\caption{\label{F:fig_mock_dust} Dust attenuated mock JWST images. We show the host galaxy of brightest quasar, plotting the distribution of dust-attenuated luminosities of star particles in NIRCam's F115W, F150W, F200W, F277W, F356W, F444W filters and MIRI's F560W, F770W filters. The images show a region of size $10 kpc$ in physical units, sampled at JWST resolution of $0.032^{"} - 0.111^{"}$, depending on the filter and include the effects of PSF. The {\em Top panel} shows the images with the contribution from AGN luminosity included. The {\em Bottom panel} images have do not have AGN.}
\end{figure*}

As we move from the left to the right panels of Figure~\ref{F:fig_mock_stellarmass}, the intensity of the pixels in the images represents the distribution of stellar mass, intrinsic UV luminosity, observed UV luminosity (i.e. with dust attenuation) and observed frame R-band luminosity respectively.
 We note that the effects of PSF and the contribution from quasar luminosity are not included here. The effects of dust attenuation are clearly visible in the middle two panels. We have already seen 
 in Figure~\ref{F:fig_sed_obs} that dust can decrease the relevant
 luminosity by a factor of $\sim 5-10$. Given that the luminosity and black hole mass of the observed quasar, J1342$+$0928 in B18 are extremely high, it is unlikely to be very attenuated. So, here we assume that the dust attenuation can be ignored for the quasar luminosity. 

\begin{figure*}
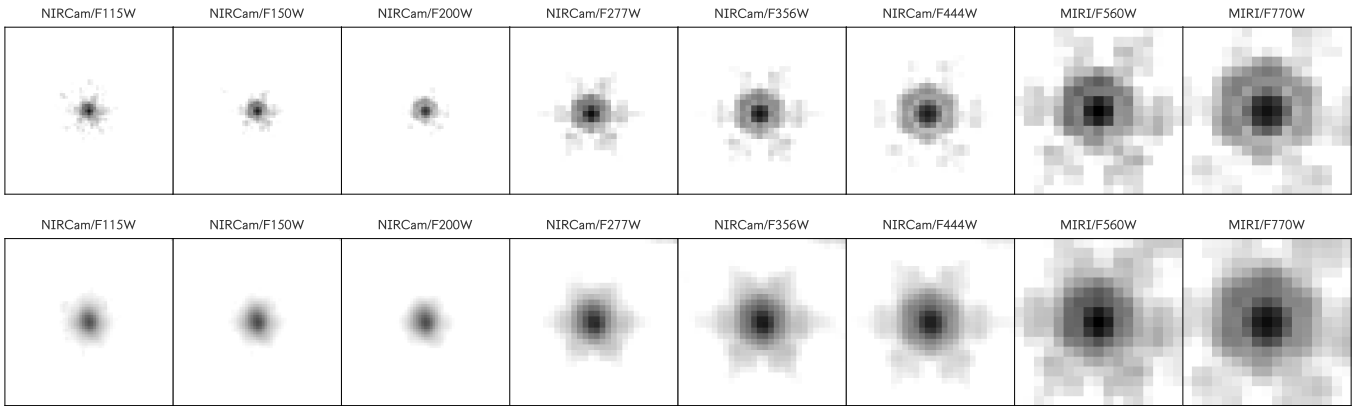

\vbox{
\includegraphics[width=2.1\columnwidth]{JWST_intrinsic_AGN_logged.pdf} \\
\includegraphics[width=2.1\columnwidth]{JWST_intrinsic_noAGN_logged.pdf}}

\caption{\label{F:fig_mock_intrinsic} Intrinsic (no dust) mock JWST images. We show  the host galaxy of brightest quasar, plotting the distribution of intrinsic luminosities of star particles in NIRCam's F115W, F150W, F200W, F277W, F356W, F444W filters and MIRI's F560W, F770W filters. The images show a region of size $10 kpc$ in physical units, sampled at JWST resolution and include the effects of PSF. The {\em Top panel} shows the images with the contribution from AGN luminosity included. The {\em Bottom} panel images do not have AGN.
}
\end{figure*}

\subsection{Mock JWST images in NIRcam and MIRI filters}

We show mock JWST images of the host galaxy of the brightest quasar in Figure~\ref{F:fig_mock_dust} and Figure~\ref{F:fig_mock_intrinsic} where the luminosity has been computed using the individual band luminosities of the star particles in the NIRCam and JWST filters.
All the mock JWST images shown in Figure~\ref{F:fig_mock_dust} and Figure~\ref{F:fig_mock_intrinsic} are of width $10 kpc$ in physical units and the PSF's for these images were generated using the WebbPSF python package \citep{2015ascl.soft04007P}.

In Figure~\ref{F:fig_mock_dust}, the luminosities were obtained from the dust attenuated SED's of the star particles. The figure shows the galaxy image in each of NIRCam filters: F115W, F150W, F200W, F277W, F356W, F444W and MIRI filters : F560W, F770W. 
In the top panel of the figure, the images show the combined luminosity from the AGN and stellar component, when taking the relevant filter-dependent PSF effects into account. These images can be compared with those in the bottom panel, where only the distribution of stellar luminosity in the galaxy is plotted after convolution with the PSF, and AGN contribution is not included. 
As expected, when the AGN is included the AGN dominates in all filters. The FWHM of the PSF is approximately in the range $0.04^{"}- 0.145^{"}$, increasing as we go to longer wavelengths for the NIRCam filters shown and $\sim 0.2^{"} - 0.25^{"}$ for the MIRI filters. Comparison of the top and bottom rows reveal that the PSF is indeed much more centrally concentrated than the galaxy light, indicating that point source subtraction should be possible in
principle. The color scale is different in the top and bottom rows (in order to make the galaxy visible in the bottom row), so that the galaxy in the top row is largely below the lower end of the scale (but it is included, together with the AGN). The elliptical shape of the galaxy in the bottom panels can be compared to that in the native images (e.g., of the stellar mass distribution) in Figure \ref{F:fig_mock_stellarmass}, indicating that if AGN subtraction is carried out it may be possible to extract this aspect of galaxy 
morphology from observations.

It is obvious from Figure ~\ref{F:fig_mock_dust} that the galaxy is extremely compact. In order to give a quantitative value for the galaxy effective radius $R_{e}$, we first exclude star particles $>10$ kpc away from the center of the galaxy. We find the exact center by minimizing the second moment of the smoothed galaxy image.  We then compute the galaxy size using a pixel based technique (summing the area of the brightest pixels that sum to to 50\% of the total light). This yields an effective radius
of $R_{E}=0.35$ kpc. This can be compared to the half-stellar mass radius, which is similar, $0.4$ kpc.
It is interesting this host galaxy is among the most compact of all galaxies with stellar mass $\sim 10^{10}  M_{\odot}$ (see Figure 4 of \citep{2015ApJ...808L..17F}) We investigate the size distribution of galaxies and the relationship to their black holes in other work (Wilkins et al., in preparation).

Figure~\ref{F:fig_mock_intrinsic} is similar to Figure~\ref{F:fig_mock_dust}, but the band luminosities of the star particles are obtained from their intrinsic SED's (i.e., without dust 
attenuation). As we have seen in the rest frame UV images in  Figure~\ref{F:fig_mock_stellarmass}, the intrinsic luminosity distribution of the galaxy is much more centrally concentrated. The dust attenuation towards the galaxy center makes a large difference to the visual impression when comparing the bottom rows of Figure~\ref{F:fig_mock_intrinsic} and Figure~\ref{F:fig_mock_dust}.
The top rows including the AGN are barely affected on the other hand, showing again that point source subtraction will be important.

\begin{figure}
\hspace{-1.5cm}
\includegraphics[width=1.25\columnwidth]{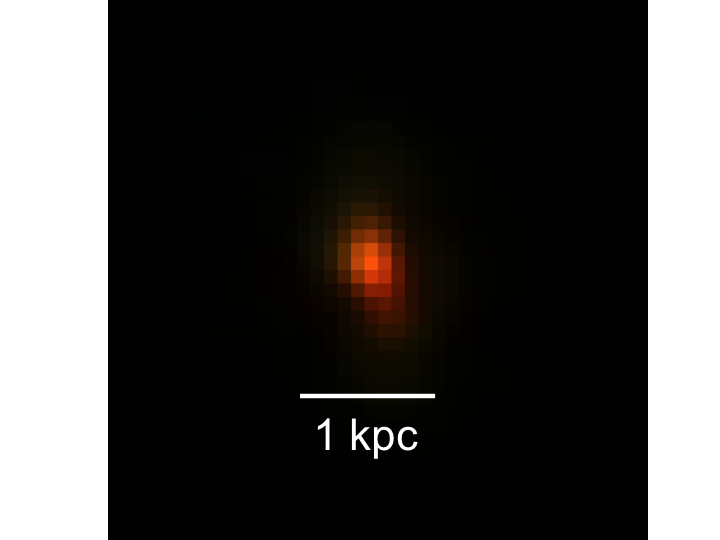}
\caption{\label{F:fig_quasar_miri}  RGB image of the host galaxy of the brightest quasar in BT-II. We use the luminosities in the K-band, V-band and B-band to form the composite image. The full width of the plot is $4$ kpc in physical units, (corresponding to an angular size of $0.8^{"}$). Each pixel is 0.1 kpc in size.} 
\end{figure}

\subsection{Composite image}

In Figure~\ref{F:fig_quasar_miri}, we show a galaxy image where the star particles are color coded by their respective luminosities in the K-band, V-band and B-band to form an RGB composite image. The pixelization reflects the pixel resolution of $0.02^{"}$, but the image does not include PSF effects. The AGN is not included. The image is fairly featureless, with small color gradients. We expect the host galaxy of the brightest quasar, J1342$+$0928 in B18 to have such an aspect when observed by JWST.

\section{Conclusions} \label{S:conclusions} 
In this paper, we studied the properties of the host galaxy of the most luminous quasar in the BlueTides-II (BT-II) simulation at redshift, $z \sim 7.5$. The first quasar observed at these redshifts was reported by \cite{2018Natur.553..473B}( hereafter B18). This quasar, J1342$+$0928 has a luminosity of $4 \times 10^{13} L_{\odot}$ and black hole mass of $8 \times 10^{8}M_{\odot}$. The volume and resolution of the BT-II simulation makes it possible to study the properties of rare objects in the early universe comparable to the B18 quasar. The brightest quasar in the simulation has a luminosity of $\sim 1.2 \times 10^{13} L_{\odot}$ and a black hole mass of $6.4 \times 10^{8}M_{\odot}$, at $z=7.6$ which is comparable to the observed quasar, J1342$+$0928 in B18. However, the host galaxy properties of the observed quasar are not completely known, except for some constraints reported by \cite{2017ApJ...851L...8V} on the host dynamical mass ($< 1.5 \times 10^{11}M_{\odot}$), star formation rate ($85-545 M_{\odot}yr^{-1}$)and dust mass ($\sim 10^{8}M_{\odot}$) using observations from IRAM/NOEMA and JVLA. JWST will make observational measurements of high red-shift galaxies including the J1342$+$0928 quasar of B18 in rest-frame optical/near-IR wavelengths. Here, in addition to reporting the properties of the host galaxy of brightest quasar in BT-II, we also make predictions for the properties of high redshift galaxies based on their AGN luminosity. Further, we compare the spectral energy distributions of the galaxy with that of the underlying quasar and show mock images of the galaxy in JWST NIRCam and MIRI filters.

The most luminous quasar in BT-II is hosted by a galaxy of stellar mass, $4 \times 10^{10} M_{\odot}$ and halo mass, $1.2 \times 10^{12} M_{\odot}$. The SFR of the galaxy is $\sim 83 M_{\odot} yr^{-1}$ at $z=7.6$ with the SFR increasing by up to a factor of $2$ as we move back to redshift $z=8$. The host galaxy is one of the more massive galaxies at this redshift. However, this galaxy is not the most massive or luminous galaxy in BT-II. There are about 10 more galaxies in BT-II that are more massive than the host of the brightest quasar. From the UV magnitudes of the galaxies in the simulation, we find that the host galaxy is fainter by about 3 magnitudes when compared to the brightest galaxies in BT-II. By comparing the mean metallicities of the galaxies and their AGN luminosity, we find that the host galaxy of the brightest quasar is among the galaxies with the highest metal content. Comparing our results with  \cite{2017ApJ...851L...8V}, we find that the predictions for stellar mass, SFR and metal enriched galaxy are consistent with the observational constraints. We also find that the galaxy is elliptical with a disk to total ($D/T$) ratio of $0.2$ and is centered in a region with low tidal field consistent with the findings in \cite{2017MNRAS.467.4243D} for galaxies hosting most massive black holes.

We computed the spectral energy distribution (SED) of the quasar and the host galaxy from the simulation data, to  compare the relative monochromatic luminosities of the quasar and it's host galaxy. Comparing the galaxy's dust-attenuated spectrum with it's intrinsic spectrum,we find that the effect of dust decreases the luminosity by a factor of $\sim 5 - 10$. The AGN is brighter than the dust-attenuated galaxy spectrum at all wavelengths and by a factor of $20-50$ times in mid and near infrared JWST bands.

Finally, we presented mock images of the host galaxy in JWST bands by taking into account the filter-dependent PSF effects and the pixelization  corresponding to each JWST filter. The BT-II simulation has a resolution of $\sim 0.05^{``}$ at $z=7.6$, while JWST has an angular resolution of $\sim 0.1^{``}$. Given that the simulation resolution is well matched to that of JWST, BT-II is ideal to make predictions for the visual appearance of the host galaxy. We also looked at images showing the stellar distribution around the quasar, covering a region of size $4\hmpc$ corresponding to JWST's field-of-view (FOV),  as well as studying the stellar distribution in the host galaxy without including PSF effects. 

The prediction from BlueTides is that there are likely to be some companion galaxies in the JWST FOV around  J1342$+$0928.  In the particular example from the simulation that we have studied there were two that were above the magnitude limit for reasonable JWST observations. As can be seen from Figure \ref{fig:Lgalaxy_Lagn_MAG}, any that are seen are unlikely to host bright AGN, and so there may an opportunity for galaxy imaging without the difficulties of
point source subtraction. Turning to the host of the brightest BT-II quasar itself,
by looking at the stellar images without PSF effects, we observe that the galaxy emission is visible up to $\sim 8 kpc$ from the quasar. The galaxy effective radius is however much less than this, $r_{E}=0.35$ kpc. The galaxy surface brightness is fairly featureless with an ellipsoidal shape, consistent with the low kinematically measured  $D/T$ ratio.  Point source subtraction of the AGN from the host galaxy images of the B18 quasar in JWST bands should be possible but will be challenging. This because the AGN outshines the host galaxy by so much and because we expect the host galaxy to be extremely compact, even though it is as massive as the Milky Way.
Follow-up HST observations of $z>6$
quasars have been elusive at revealing the underlying UV stellar light of their host galaxies \citep[e.g.,][]{{2012ApJ...756..150D}, {2012ApJ...756L..38M}}. JWST's exquisite sensitivity, resolution and wide wavelength coverage will be
essential (and hopefully sufficient) to constrain the stellar mass of these tiny host galaxies.

\section*{Acknowledgements}
We acknowledge funding from NSF
ACI-1614853, NSF AST-1517593, NSF AST-1616168, NASA ATP NNX17AK56G and NASA ATP 17-0123 and the BlueWaters PAID program. The \texttt{BLUETIDES} simulation was run on the BlueWaters facility at the National Center for Supercomputing Applications

\bibliographystyle{mnras} \bibliography{draft}
\end{document}